# Spin-Hall and Anisotropic Magnetoresistance in Ferrimagnetic Co-Gd / Pt layers

W. Zhou,[1,†] T. Seki,[1,2,*] T. Kubota,[1,2] G. E. W. Bauer[1-4], and K. Takanashi,[1,2]

[1]*Institute for Materials Research, Tohoku University, Sendai 980-8577, Japan*

[2]*Center for Spintronics Research Network, Tohoku University, Sendai 980-8577, Japan*

[3]*AIMR, Tohoku University, Sendai 980-8577, Japan*

[4]*Zernike Institute for Advanced Materials, University of Groningen, The Netherlands*

\* e-mail: go-sai@imr.tohoku.ac.jp

[†]Present affiliation: National Institute for Materials Science, Tsukuba, Japan

**Abstract:**

We present the Co-Gd composition dependence of the spin-Hall magnetoresistance (SMR) and anisotropic magnetoresistance (AMR) for ferrimagnetic $Co_{100-x}Gd_x$ / Pt bilayers. With Gd concentration $x$, its magnetic moment increasingly competes with the Co moment in the net magnetization. We find a nearly compensated ferrimagnetic state at $x = 24$. The AMR changes sign from positive to negative with increasing $x$, vanishing near the magnetization compensation. On the other hand, the SMR does not vary significantly even where the AMR vanishes. These experimental results indicate that very different scattering mechanisms are responsible for AMR and SMR. We discuss a possible origin for the alloy composition dependence.

# 1. Introduction

Antiferromagnetic spintronics [1-6] is an emerging research field that has attracted much attention because of the unique properties of antiferromagnets: zero net magnetization, small magnetic susceptibility [7], and magnetization dynamics characteristically different from ferromagnets [8-11]. Antiferromagnets have great potential for the development of novel spintronic devices such as crosstalk-free and ultrahigh-density non-volatile memories because they do not generate and are robust against magnetic stray fields [12]. However, several problems have to be solved before exploiting the aforementioned functionalities in practical devices. A major issue is the efficient control of the antiferromagnetic order. The small magnetic susceptibility of antiferromagnets renders magnetic field control difficult. Current-induced spin transfer phenomena may be a possible solution for this problem [13,14]. Recent studies demonstrate that spin current ($J_s$) can be generated by an antiferromagnet [4,15] and also interacts with its magnetic moments [2,3,5,11,14,16-19]. However, the coupling phenomenon between antiferromagnetic order and spin currents has not yet been fully understood. It is a complex problem involving the spin-dependent scattering in the bulk and at interfaces to electric contacts. Here we focus on the latter by revealing details of the spin mixing at the interface between platinum and a ferrimagnet around the compensation point.

Co-Gd amorphous alloys are ferrimagnets, in which the Co and Gd moments ($m^{Co}$ and $m^{Gd}$) are coupled antiferromagnetically [20]. The net magnetic moment of Co-Gd ($m^{Co-Gd}$) is given by $| m^{Co} - m^{Gd} |$, meaning that dominance of one of them in the $m^{\text{Co-Gd}}$ magnetization strongly depends on the alloy composition as shown in **Figs. 1(a) and 1(b)**. The ferrimagnetic state with zero magnetization at the Co-Gd compensation point resembles an antiferromagnet. By exploiting ferrimagnetic materials such as Co-Gd and Co-Fe-Gd, several studies recently reported an interaction between $J_s$ and magnetizations near the compensation points. Ham *et al.* [21] found an enhanced damping-like component of the spin-orbit torque (SOT) near the compensated perpendicularly magnetized $Gd_{25}Fe_{65.6}Co_{9.4}$ / Pt. This was explained by the reduction of the net magnetic moment. Although Mishra *et al.* [22] also observed a substantial increase of the SOT effective field and switching efficiency in perpendicularly magnetized Co-Gd / Pt, they conclude that the negative exchange interaction in the ferrimagnet enhances the SOT near compensation. Co-Gd alloys can also display angular momentum compensation, which is beneficial for ultra-fast magnetization dynamics: Kim *et al.* [23] demonstrated magnetic field-driven fast domain wall motion in a ferrimagnetic Co-Fe-Gd wire at the angular momentum compensation temperature. Even though Co-Gd is an important material class, both from fundamental and application point of view, the detailed mechanism of spin-dependent transport in this material is not well understood. Here we present a systematic study of magnetotransport of Co-Gd alloy / Pt thin films that accesses spin-dependent scattering parameters and sheds light on the interaction between $J_s$ and the ferrimagnetic order. Our analysis separates the contributions from the spin-Hall magnetoresistance (SMR) and the anisotropic magnetoresistance (AMR) that occur simultaneously in all-metal magnetic bilayers, which should help to establish microscopic models for both effects.

We report the different composition dependences of SMR and AMR for the Co-Gd / Pt bilayers with in-plane magnetization. The sign of the AMR monotonically changes from positive to negative by increasing the Gd concentration and vanishes near the magnetization compensation composition. On the other hand, the SMR remains finite even when the AMR vanishes, which is a direct proof for different physics. We interpret the composition dependence of the SMR in terms of a spin mixing conductance that, in contrast to the conventional wisdom, depends on the magnet.

**2. Experimental Procedure**

Thin films were deposited on a thermally oxidized Si substrate using an ultrahigh vacuum compatible magnetron sputtering system with the base pressure below $2 \times 10^{-7}$ Pa. First, a 4 nm-thick Cr buffer was deposited on the Si-O substrate. Then Co and Gd were co-deposited to form the $Co_{100-x}Gd_x$ layers with a thickness of 30 nm. Finally, a 4 nm-thick Pt layer was deposited. All the layers were deposited at room temperature. By tuning the sputtering powers of Co and Gd targets, the Gd concentration $x$ (at. %) was widely varied from $x = 0$ to $x = 45$. Except for $x = 0$, *i.e.* pure cobalt, the Co-Gd layers were amorphous alloys, as confirmed by reflection high energy electron diffraction (RHEED) in **Fig. 1(c)**. In contrast to the amorphous phase of Co-Gd, the RHEED pattern of **Fig. 1(d)** indicates that the top Pt layer crystallizes on the amorphous Co-Gd. The top Pt layer serves as not only the capping layer to prevent the Co-Gd from oxidation, but also as converter of a charge current ($J_c$) to a transverse spin current $J_s$ by the spin-Hall effect (SHE) [25]. We also prepared reference samples consisting of Al / $Co_{100-x}Gd_x$ / Al, for which we anticipated negligible SMR because of the small spin-orbit coupling in Al. Magnetic properties were measured by a superconducting quantum interference magnetometer, a vibrating sample magnetometer and a longitudinal magneto-optical Kerr effect (L-MOKE) set-up with laser wavelength of 680 nm, all at room temperature.

The thin films were patterned into a 10 μm-wide Hall-cross by photolithography and Ar ion milling. In order to separate the contributions of AMR and SMR as depicted in **Fig. 2**, we measured the magnetoresistance as a function of direction of an applied magnetic field (**H**) in two configurations. In the $\gamma$ scan, **H** rotates in the $x$ - $z$ plane (**Fig. 2(a)**), anticipating an AMR since $J_c$ flows along the $x$ direction. The SMR is accessed by the $\beta$ scan in which **H** rotates in the $y$ - $z$ plane (**Fig. 2(b)**) [26]. In these measurements we apply a large magnetic field |**H**|=70 kOe such that the magnetization and field are collinear to a good approximation. All measurements were carried out at room temperature.

## 3. Experimental Results and Discussion

### A. Composition Dependence of the Magnetic Properties

**Figs. 3(a) - 3(d)** show the magnetization versus magnetic field ($M$ - $H$ curves) of Cr / Co$_{100-x}$Gd$_x$ / Pt with (a) $x$ = 12, (b) 24, (c) 25 and (d) 37 and **Figs. 3(e) - 3(h)** the corresponding L-MOKE loops. **H** was applied in the film plane. When $x$ is increased from 12 to 37, the net magnetization is suppressed at $x$ = 24 and 25 and accompanied by an increased coercivity ($H_c$). $M$ is observed to increase again for $x$ = 37. In contrast to the $M$ - $H$ curves, the L-MOKE loops show a gradual decrease in the magnitude of Kerr rotation angle ($\theta_K$) with $x$. A remarkable feature of the L-MOKE loops is the sign reversal of $\theta_K$ between $x$ = 24 and 25.

The solid circles in **Fig. 4** summarize (a) $M$, (b) the maximum $\theta_K$ and (c) its absolute value as a function of $x$ for Cr / Co$_{100-x}$Gd$_x$ / Pt. As $x$ is increased, $M$ shows a local minimum at $x$ = 24 and $\theta_K$ reverses its sign at $x$ = 25. $|\theta_K|$ slowly and monotonically decreases with $x$. We conclude that the room-temperature magnetization compensation point of Co-Gd lies between $x$ = 24 and 25. This composition is close to the reported magnetization compensation point at room temperature [20]. We attribute the sign reversal of $\theta_K$ to the change of the dominant magnetic component from $m^{Co}$ to $m^{Gd}$. The present L-MOKE system is equipped with a 680-nm-wavelength semiconductor laser, which selectively probes $m^{Co}$. At the Co-rich composition ($x$ = 12), $m^{Co}$ dominates and is parallel to **H**, resulting in a positive $\theta_K$. On the other hand, when $m^{Gd}$ dominates, $m^{Co}$ is antiparallel to **H** and $\theta_K$ is negative. The results for the Al / Co$_{100-x}$Gd$_x$ / Al reference samples, denoted by open squares in **Fig. 4**, exhibited magnetic properties very similar to Cr / Co$_{100-x}$Gd$_x$ / Pt.

*B. Field Angular Dependence of SMR and AMR*

Here we present results of the high field (70 kOe) SMR and AMR measurements for Cr / $Co_{100-x}Gd_x$ / Pt. The $\gamma$ scans of the longitudinal resistance ($R_{xx}$) are displayed in **Figs. 5(a) - 5(d)** ((a) $x$ = 12, (b) 24, (c) 25 and (d) 45), and the corresponding $\beta$ scans in **Figs. 5(e) - 5(h)**. $J_c$ was set at 0.1 mA or current density of $2.6 \times 10^4$ A $cm^{-2}$. As shown in **Fig. 2**, the magnetization of the $\gamma$ ($\beta$) scan lies in the $x$ - $z$ ($y$ – $z$) plane, which corresponds to the AMR (SMR), respectively. The AMR ratio $AMR = \left\{ \left( R_{xx}^{\gamma=90°} - R_{xx}^{\gamma=0°} \right) / R_{xx}^{\gamma=0°} \right\} \times 100 = \left\{ \left( \rho_{//} - \rho_{\perp} \right) / \rho_{\perp} \right\} \times 100$ in terms of the longitudinal ($\rho_{//}$) and transverse ($\rho_{\perp}$) resistivities. The definition of $SMR = \left\{ \left( R_{xx}^{\beta=90°} - R_{xx}^{\beta=0°} \right) / R_{xx}^{\beta=0°} \right\} \times 100 = \left\{ \Delta\rho / \rho_0 \right\} \times 100$, where $\rho_0$ is the resistivity at the SMR maximum and $\Delta\rho$ the resistivity modulation. This definition agrees with previous ones for metallic bilayers [27], but differs from that used for magnetic insulators / Pt [26]. Therefore, a negative SMR here corresponds to the "normal" situation in Ref. [28]. The longitudinal electric fields along $J_{c,x}$ due to AMR ($E_{xx}^{AMR}$) and SMR ($E_{xx}^{SMR}$) are well described by [26, 27]:

$$E_{xx}^{AMR} = \left\{ \rho_{\perp} + \left( \rho_{//} - \rho_{\perp} \right) \sin^2 \gamma \right\} J_x, \quad (1)$$

and

$$E_{xx}^{SMR} = \left\{ \rho_0 + \Delta\rho \sin^2 \beta \right\} J_x. \quad (2)$$

The solid lines in **Fig. 5** denote fits by Eqs. (1) and (2), which is a strong evidence that the angular dependences indeed are caused by AMR and SMR.

AMR is the dependence of the resistance on the angle between current and magnetization and defined to be positive when $\rho_{//} > \rho_\perp$. The SMR is generated by the spin-orbit interaction in the normal metal layer and the exchange interaction at the interface. Here, we find for Gd-Co alloys an AMR > 0 for $x$ = 12, while AMR < 0 for $x$ = 45, and AMR = 0 at $x$ = 25. In contrast, the SMR ratio is negative regardless of $x$ even at the composition for which the AMR vanishes. As mentioned above, a *negative* sign of SMR in the convention of Ref. [27] implies a net magnetization of Co-Gd is parallel to the external magnetic field.

**Figure 6** summarizes (a) the composition dependence of the longitudinal resistance $R$ (on this scale the dependence on magnetization direction is negligibly small), (b) AMR ratio and (c) SMR ratio for Cr / $Co_{100-x}Gd_x$ / Pt. The experimental values of $R$, AMR ratio and SMR ratio are also summarized in **Table 1**. Compared with pure Co, the alloy scattering of amorphous Co-Gd strongly increases the resistivity. **Figure 6(b)** clearly demonstrates the sign change of the AMR from positive to negative as $x$ increases (see the inset of **Fig. 6(b)**). Pure Co shows a positive AMR [29] while a negative AMR has been reported for a Gd single crystal [30]. Our results suggest that the *d*-electrons of Co and the *f*-electrons of Gd contribute oppositely to the AMR phenomenon. Hence, the effect of local *s-d* (*s-f*) scattering appears to cancel (to the experimental accuracy) exactly at the compensation point. However, the microscopic mechanism of the AMR is much more complicated, being governed by the full electronic structure, see e.g. Ref. [31]. Nevertheless, the vanishing of the AMR at the compensation point is most likely not a coincidence and our results should help to develop better theoretical models for spin and charge transport in magnetic metals. While the AMR changes sign at $x = 25$, the SMR is negative regardless of $x$, which implies that the scattering mechanisms of AMR and SMR are very different.

In order to shed light on this matter we carried out $\gamma$ and $\beta$ scan magnetoresistance measurement for the Al / Co$_{100-x}$Gd$_x$ / Al reference samples as shown in **Fig. 7**. The Co-rich Co-Gd with $x = 7$ (**Fig. 7a**) exhibits a positive AMR whereas in the Gd-rich Co-Gd with $x = 42$ (**Fig. 7b**) AMR < 0. This sign change is consistent with the results for Cr / Co$_{100-x}$Gd$_x$ / Pt, and proves that the AMR is qualitatively not affected by the normal metal and the interfaces. The absence of a clear SMR for the reference samples in **Figs. 7c and 7d** can be attributed to the small spin-orbit coupling and negligibly small spin Hall effect in Al as anticipated. We therefore may conclude that (i) the sign change in the $\gamma$ scan with increasing $x$ originates from the bulk scattering in the Co-Gd layer, and (ii) the SMR for the Cr / Co$_{100-x}$Gd$_x$ / Pt is dominantly caused by the direct and inverse spin Hall effects without a significant bulk contribution. A contribution from the transverse AMR [32] to the $\beta$ scans of $R_{xx}$ can be excluded because the Al / Co$_{100-x}$Gd$_x$ / Al sample resistance does not change in the $\beta$ scans.

*C. Discussion*

We now discuss the composition dependence of SMR for the Cr / Co$_{100-x}$Gd$_x$ / Pt. In metallic bilayers of a nonmagnet (N) with large spin-orbit coupling and a ferromagnet (F) [27]

$$\frac{\Delta\rho}{\rho_0} \sim -(\theta_{SH})^2 \frac{\lambda_N}{t_N} \frac{\tanh^2(t_N/2\lambda_N)}{1+\xi} \left\{ \frac{g_R}{1+g_R \coth(t_N/\lambda_N)} - \frac{g_F}{1+g_F \coth(t_N/\lambda_N)} \right\}, \quad (3)$$

in terms of

$$\xi \equiv \frac{\rho_N t_F}{\rho_F t_N}, \quad (4)$$

$$g_R \equiv 2\rho_N \lambda_N \operatorname{Re}[G_{MIX}], \quad (5)$$

$$g_F \equiv \frac{(1-P^2)\rho_N \lambda_N}{\rho_F \lambda_F \coth(t_F/\lambda_F)}, \quad (6)$$

where $\rho_{N(F)}$, $\lambda_{N(F)}$, $t_{N(F)}$ are the resistivity, spin diffusion length, thickness of the N (F) layer, respectively, $\theta_{SH}$ is the spin-Hall angle of the N layer, $P$ is the current spin polarization of the F layer, and $G_{MIX}$ is the spin mixing conductance of the interface. The first term in the curly bracket of Eq. (3) coincides with the expression for the SMR for a ferromagnetic insulator. The second term takes the absorption of the longitudinal spin current by the ferromagnetic metal into account (and is not to be confused with the unidirectional SMR [33]).

Calculating the SMR ratio Eq. (3), *i.e.* $(\Delta\rho/\rho_0) \times 100$, requires values of many parameters. The Gd concentration dependence of $\rho_F$ is displayed in **Fig. 8a**. $\rho_N$ for Pt was experimentally measured using the 4 nm-thick Pt single layer film, and was obtained to be $4 \times 10^{-7}$ Ω m. We cannot simply measure $\lambda_F$ in our Co-Gd, but it is a few nm at most and we assume $\lambda_F \approx 2$ nm in the following. $P = 0.3$ has been derived from the tunneling spin polarization of Co-Gd [34]. $G_{MIX}$ is a measure of the transverse spin current absorption that we initially choose to not depend on $x$, *e.g.* $G_{MIX} = 1 \times 10^{15}$ Ω$^{-1}$ m$^{-2}$. $\theta_{SH} \approx 0.08$ and $\lambda_N \approx 3$ nm reported for the Pt [Ref. 35] are chosen. **Figure 8b** compares the observed absolute values of the experimental and model SMR ratios for each alloy composition, where the small scatter of calculated SMR reflects that in the measured resistivities. Those agree quite well, but in contrast to the experimental observation, the absolute value of the calculated SMR ratios increases with increasing $x$. This calculated tendency as a function of composition does not agree with the experimental trend, suggesting an alloy concentration-dependence of material parameters that we assumed constant, such as $\lambda_F$ and $G_{MIX}$. $P$ affects the SMR because a spin current can penetrate the metallic ferromagnets when polarized parallel the magnetization. However, because the alloy resistance is relatively high, the calculated SMR does not change much for $P \leq 0.4$.

We can turn the table and calculate the parameter dependence on Gd concentration. The dependence would reproduce the experiments. Here, we focus on $G_{MIX}$. The Gd concentration dependence of $G_{MIX}$ that results from inverting Eq. (3) is shown in **Fig. 8c**. $G_{MIX}$ is seen to decrease strongly with increasing $x$, implying that the ratio of $m^{Co}$ versus $m^{Gb}$ at the interface to Pt plays an important role in the spin mixing. This is surprising, since theory predicts that spin mixing is mainly governed by the electron density [36] or the dynamical spin susceptibility [37] of the normal metal. However, these theories do not take the spin-orbit interaction into account, which might importantly modify the spin-mixing conductance of the CoGd / interface and cause its suppression with increasing Gd concentration. M. A. Schoen et al. [38] report a compositional dependence of $G_{MIX}$ for the $Ni_xCo_{1-x}$, $Ni_xFe_{1-x}$, and $Co_xFe_{1-x}$, even for such $3d$ transition-metal binary alloys. M. Tokaç et al. [39] report that $G_{MIX}$ depends on the crystal structure even for elemental Co.

*D. In-plane Field Angular Dependence*

The dependence of $R_{xx}$ on the magnetic field angle $\alpha$ (left panel in **Fig. 9**) is plotted in **Fig. 9** for $x = 25$ at which the AMR vanishes. The $\alpha$ scan should therefore give identical results with the $\beta$ scan, as confirmed by comparison with **Fig. 5(g)**. In a Co-Gd alloy with $x = 25$ the AMR and associated planar Hall effect vanish, which could be a technical advantage. For example, in attempts to measure the spin Hall angle by the spin pumping technique, spurious contribution from the planar Hall effect must be subtracted [40]. This technical difficulty can be overcome by choosing Co-Gd (or in fact any other ferrimagnet) at its compensation composition or temperature to accurately measure the spin-Hall effect, even in the case of metallic bilayer system.

**4. Summary**

We systematically investigated the Co-Gd composition dependence of the SMR and AMR of in-plane magnetized $Co_{100-x}Gd_x$ / Pt layers. As $x$ increases, the dominant magnetization changes from the $m^{Co}$ to the $m^{Gd}$ sublattices. We realized a nearly compensated ferrimagnetic structure at $x = 24$ (at room temperature). We find distinctly different composition dependences of the SMR and AMR. The AMR decreases monotonically with increasing $x$ and changes sign near the compensation composition while SMR remain constant when the AMR sign chances. The composition dependence of the AMR suggests a local picture of the AMR in which the magnetic moments of the Co and Gd contribute with opposite sign with canceling contributions at the compensation. The observed composition dependence of the SMR can be explained by an exchange interaction of the conduction electrons in Pt that is dominated by the Co magnetic moments or the spin-orbit interaction at the interface. Our findings contribute to a better understanding of an important material class.

**Acknowledgement**

The authors thank J. Barker and S. Takahashi for valuable discussions. Y. Murakami and I. Narita provided technical support during the structural characterization. This work was supported by the Grant-in-Aid for Scientific Research B (16H04487) and Innovative Area "Nano Spin Conversion Science" (26103006) as well as the Research Grant from the TEPCO Memorial Foundation. The device fabrication was partly carried out at the Cooperative Research and Development Center for Advanced Materials, IMR, Tohoku University.

**Table 1** Experimental values of *R*, AMR ratio and SMR ratio for Cr / Co$_{100-x}$Gd$_x$ / Pt.

| *x* (at. %) | *R* (Ω) | AMR ratio (%) | SMR ratio (%) |
|:---:|:---:|:---:|:---:|
| 0 | 24.1 ± 0.1 | 1.5 ± 0.002 | -0.200 ± 0.001 |
| 12 | 202.5 ± 0.1 | 0.050 ± 0.001 | -0.045 ± 0.003 |
| 19 | 257.8 ± 0.7 | 0.022 ± 0.002 | -0.043 ± 0.001 |
| 24 | 257.9 ± 0.1 | 0.012 ± 0.002 | -0.035 ± 0.001 |
| 25 | 272.3 ± 0.6 | 0 ± 0.002 | -0.036 ± 0.001 |
| 30 | 280.7 ± 3.3 | 0 ± 0.002 | -0.020 ± 0.001 |
| 34 | 272.8 ± 0.3 | -0.003 ± 0.002 | -0.030 ± 0.002 |
| 37 | 274.6 ± 0.1 | -0.007 ± 0.001 | -0.025 ± 0.002 |
| 39 | 295.3 ± 0.1 | -0.011 ± 0.003 | -0.020 ± 0.001 |
| 45 | 301.0 ± 0.1 | -0.012 ± 0.001 | -0.019 ± 0.004 |

**Figure Captions**

**Figure 1** (a,b) Relationship between net magnetization, Co magnetic moment, and Gd magnetic moment for (a) Co-rich Co-Gd and (b) Gd-rich Co-Gd. (c,d) Reflection high-energy electron diffraction patterns of the (c) Co-Gd surface ($x$ = 12) and (d) Pt surface.

**Figure 2** Measurement configurations for the angular dependence of (a) anisotropic magnetoresistance (AMR) and (b) spin-Hall magnetoresistance (SMR) of the $Co_{100-x}Gd_x$ / Pt bilayer. A charge current ($J_c$) was applied along the $x$ direction. The AMR is observed when the external magnetic field (**H**) is rotated in the $x$-$z$ plane by the angle of $\gamma$. The SMR is the resistance change when **H** is rotated in the $y$-$z$ plane by an angle $\beta$.

**Figure 3** Magnetization curves for the $Co_{100-x}Gd_x$ / Pt bilayer films with (a) $x$ = 12, (b) 24, (c) 25 and (d) 37, at room temperature and magnetic field **H** in the film plane. The longitudinal magneto-optical Kerr effect (L-MOKE) loops for (e) $x$ = 12, (f) 24, (g) 25 and (h) 37.

**Figure 4** (a) Gd concentration ($x$) dependence of magnetization ($M$), (b) the maximum Kerr rotation angle ($\theta_K$), and (c) the absolute value of $\theta_K$ ($|\theta_K|$). The solid circles represent the data for Cr / $Co_{100-x}Gd_x$ / Pt while the open squares are those of the reference samples Al / $Co_{100-x}Gd_x$ / Al.

**Figure 5** Angular ($\gamma$) dependence of the AMR of the Cr / Co$_{100-x}$Gd$_x$ / Pt with (a) $x$ = 12, (b) 24, (c) 25 and (d) 45, and angular ($\beta$) dependence of SMR for the Cr / Co$_{100-x}$Gd$_x$ / Pt with (e) $x$ = 12, (f) 24, (g) 25 and (h) 45. $J_c$ was set at 0.1 mA, and $H$ = 70 kOe was applied. The solid curves are fits by Eqs. (1) and (2).

**Figure 6** (a) Gd concentration ($x$) dependence of device resistance ($R$), (b) AMR and (c) SMR for the Cr / Co$_{100-x}$Gd$_x$ / Pt. The inset of (b) is a magnified plot of the AMR versus $x$.

**Figure 7** Angular ($\gamma$) dependence of AMR of the reference samples Al / Co$_{100-x}$Gd$_x$ / Al with (a) $x$ = 7 and (b) 42, and angular ($\beta$) dependence of SMR for the Al / Co$_{100-x}$Gd$_x$ / Al with (c) $x$ = 7 and (d) 42. $J_c$ was set to 0.1 mA, and $H$ = 70 kOe was applied. The solid curves are fits by Eq. (1).

**Figure 8** (a) Gd concentration ($x$) dependence of resistivity ($\rho_F$) of the present Co-Gd. (b) Comparison of the absolute values of the SMR ratios between the experiment (solid circles) and the model (solid line). (c) $G_{\text{MIX}}$ calculated from the SMR ratio as a function of $x$ for the Cr / Co$_{100-x}$Gd$_x$ / Pt by inverting Eq. (3).

**Figure 9** Illustration of the in-plane field angular ($\alpha$) dependence of magnetoresistance, and $\alpha$ scan for the Cr /

$Co_{100-x}Gd_x$ / Pt with $x = 25$. $J_c$ was set at 0.1 mA, and $H = 70$ kOe was applied. The solid curve is the fit by Eq. (2) with an offset of 90 degree for $\alpha$.

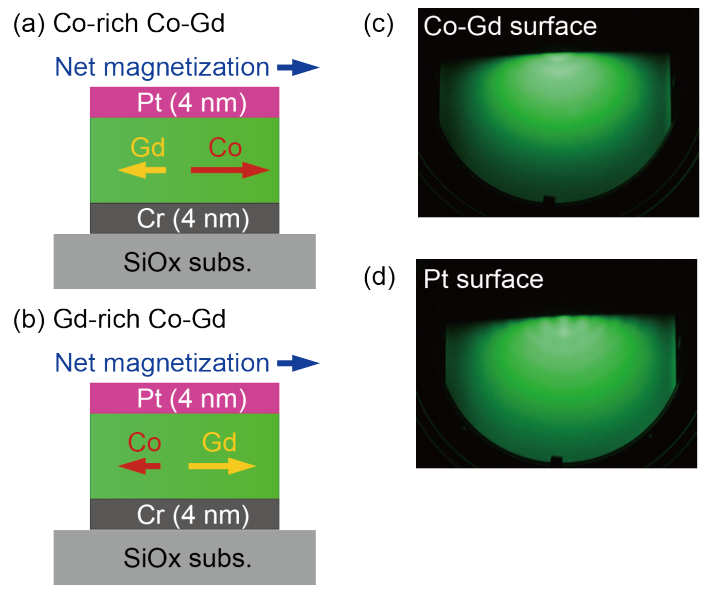

Figure 1

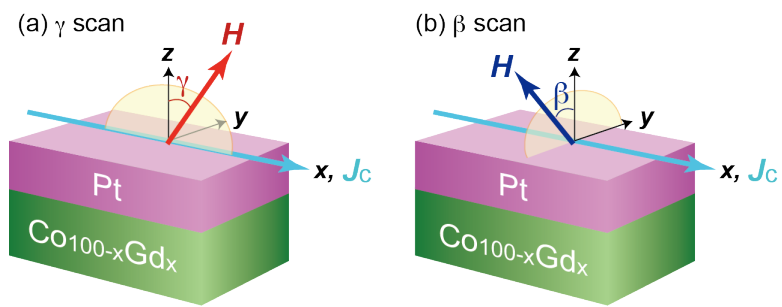

Figure 2

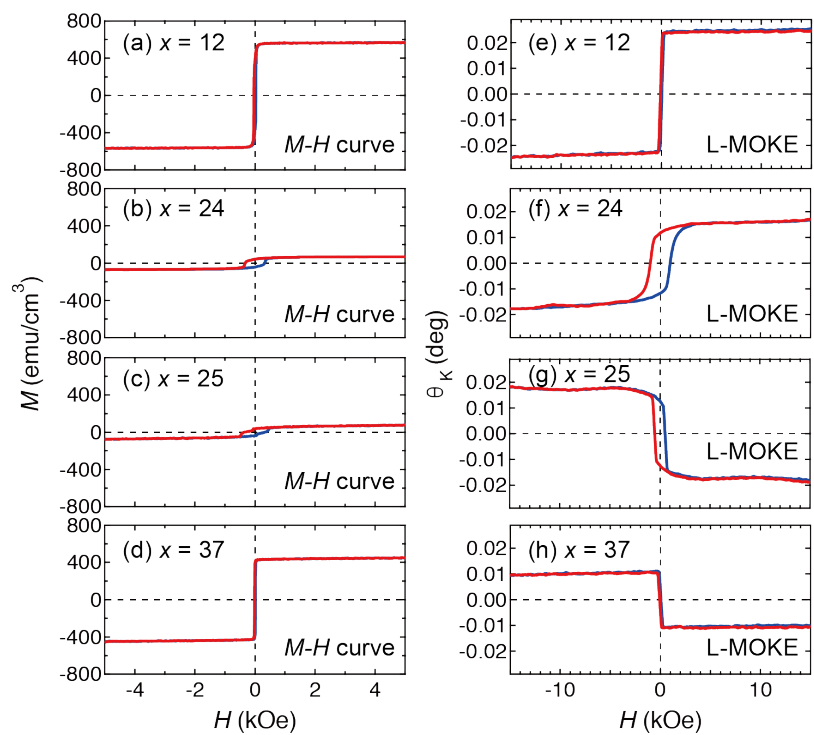

Figure 3

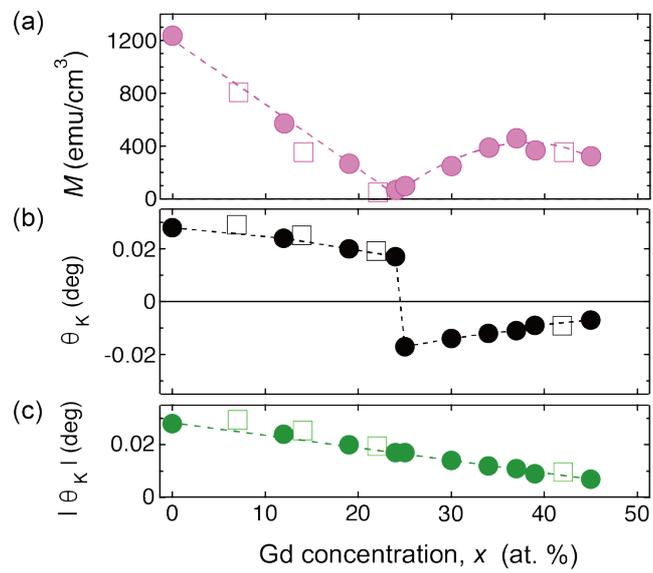

Figure 4

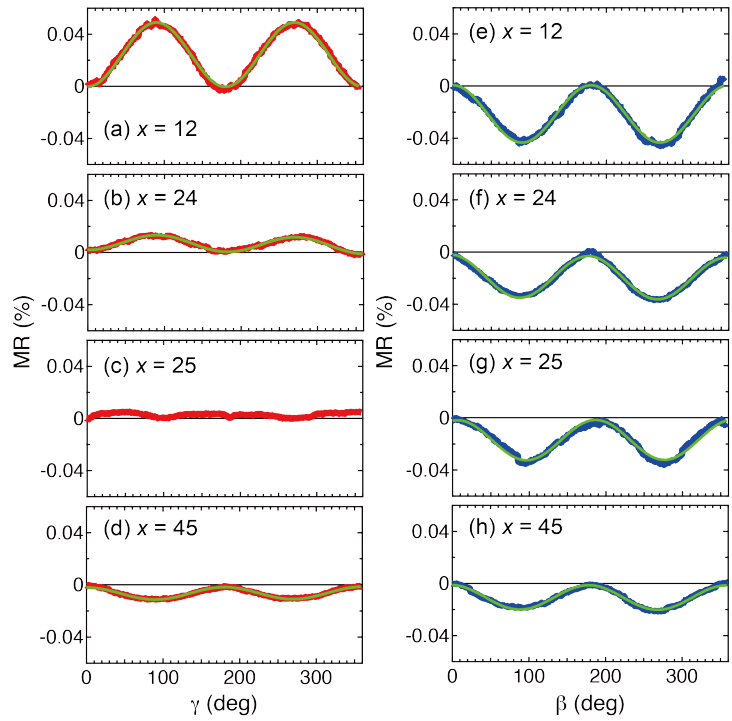

Figure 5

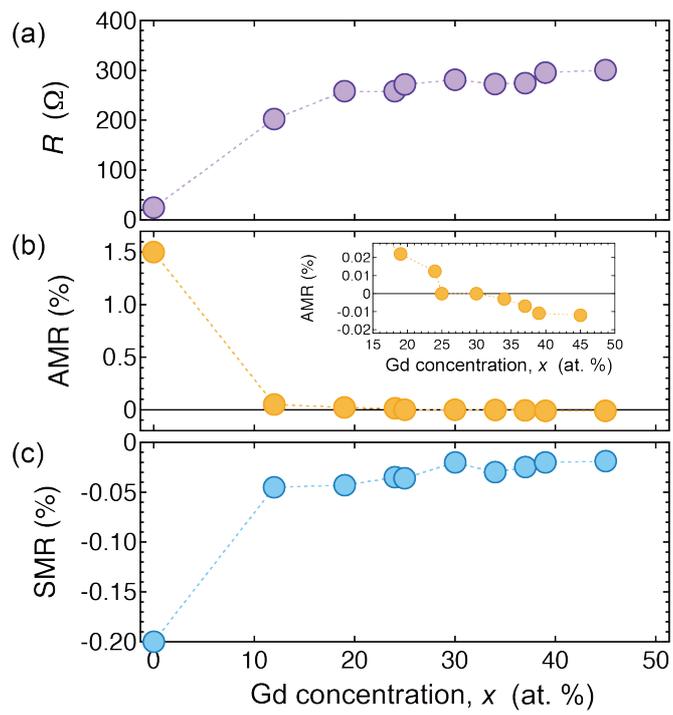

Figure 6

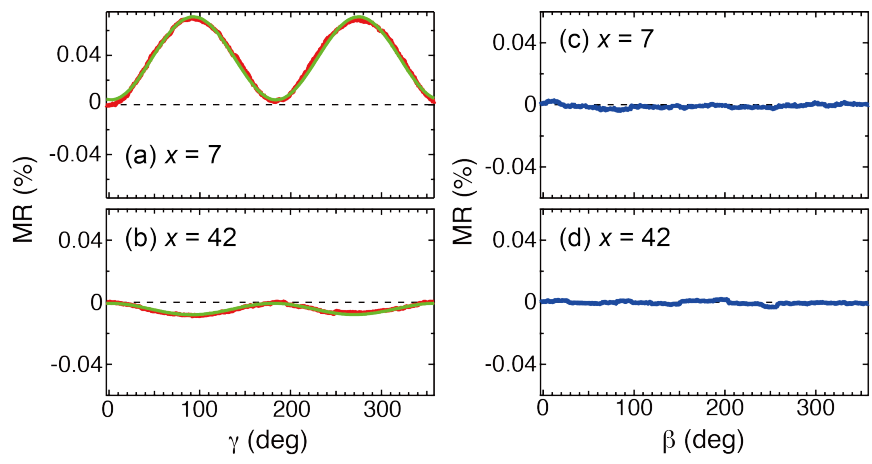

Figure 7

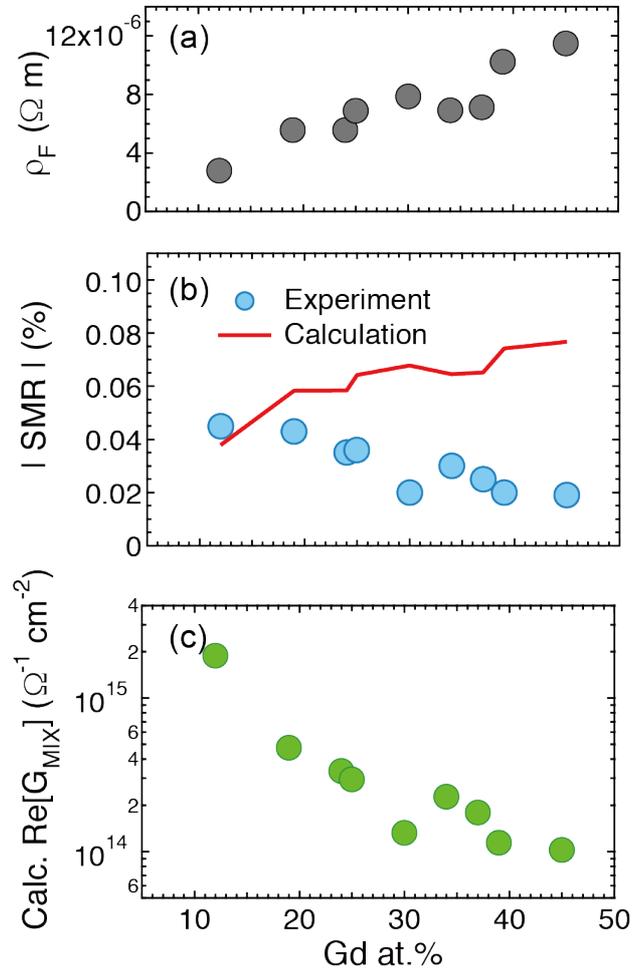

Figure 8

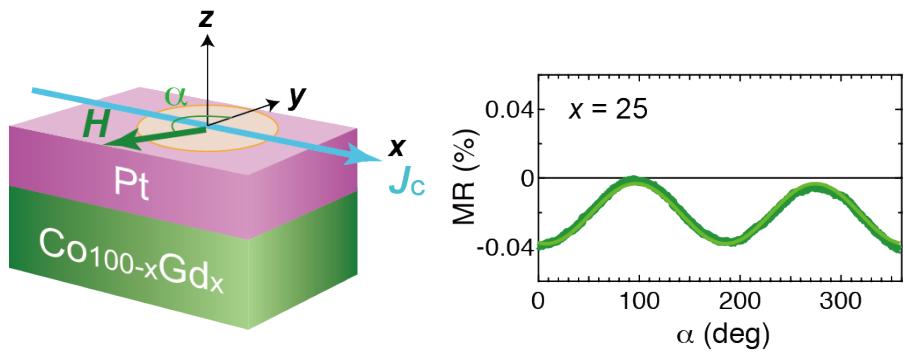

Figure 9